\documentclass[a4paper,10pt,twoside]{cpc-hepnp}

\usepackage{multicol}
\usepackage{graphicx}
\usepackage{booktabs}
\usepackage{amssymb,bm,mathrsfs,bbm,amscd}
\usepackage[tbtags]{amsmath}
\usepackage{lastpage}
\usepackage{epsfig}
\usepackage{epstopdf}

\begin{document}

%\fancyhead[co]{\footnotesize F. Jone et al: Instruction for typesetting manuscripts}

\footnotetext[0]{Received 25 November 2009}

\title{QGP tomography with photon tagged jets in ALICE\thanks{This work is supported partly by the NSFC (10875051 and 10635020),
the State Key Development Program of Basic Research of China
(2008CB317106), the Key Project of Chinese Ministry of Education
(306022 and IRT0624) and the Programme of Introducing Talents of
Discipline to Universitiesnder of China: B08033}}

\author{%
      Yaxian Mao$^{1,3}$,\email{maoyx@iopp.ccnu.edu.cn}%
\quad Yves Schutz$^{2}$,%
\quad Daicui Zhou$^{1}$,%
\quad Christophe Furget$^{3}$,%
\quad Gustavo Conesa Balbastre$^{4}$
}
\maketitle

\address{%
1~(Institute of Particle Physics, Huazhong Normal University, Wuhan 430079, China )\\
2~(CERN, Geneva 23, Switzerland)\\
3~(Laboratoire de Physique Subatomique et de
Cosmologie, CNRS/IN2P3, Grenoble 38026, France)\\
4~(Laboratori Nazionali di Frascati, INFN, Frascati, Italy) 
}

\begin{abstract}
$\gamma +$jet events provide 
%a unique mean to study medium modified jet
%properties and therefore provides 
a tomographic measurement of the
medium formed in heavy ion collisions at LHC energies. Tagging events with a well identified high $p_{T}$ direct photon and
measuring the correlation distribution of hadrons emitted oppositely to the
photon,  allows us to determine, with a good approximation, both
the jet fragmentation function and the back-to-back azimuthal
misalignement of the direct photon and the jet. Comparing
these two observables measured in $pp$
collisions with the ones measured in $AA$ collisions will reveal
the modifications of the jet structure induced by the medium
formed in $AA$ collisions and consequently will infer the medium
properties. 
\end{abstract}

\begin{keyword}
direct photon, QGP, jet structure, tomography, path length 
\end{keyword}

\begin{pacs}
24.85.+p, 25.75.Bh, 25.75.Cj, 25.75.Nq
\end{pacs}

\begin{multicols}{2}

\section{Introduction}
The Large Hadron Collider (LHC) at CERN, will collide heavy\-ions at unprecedented high energies, exceeding by a factor 30 the energy available at RHIC~\cite{rhic}. The main objective of ALICE (A Large Ion Collider Experiment)~\cite{alice}, is to study matter under extreme conditions of energy density to gain a better understanding of the fundamental properties of the strong interaction. In particular, ALICE will explore the Quark-Gluon Plasma (QGP), the state of deconfined matter predicted by QCD~\cite{QCD}.  The medium formed in heavy-ion collisions can be best probed by hard scattered partons produced in 2$\rightarrow$2 QCD processes at the leading order (LO) including in the final state a hard direct photon (Compton scattering: q + g $ \rightarrow  \gamma $+q and quark annihilation: q+$ \bar{q}  \rightarrow  \gamma$+g) . On one hand, the 4-momentum of the scattered parton is modified while traversing the medium, and on the other hand, the scattered photon does not interact, thus providing a reference for the 4-momentum of the partner parton. Hence, from the modification experienced by the hard scattered partons, measured though photon tagged jets, the medium properties can be inferred. In particular, since these hard scattering processes
%jet observable  is strongly biased towards surface production, whereas photons 
sample the entire collision volume, the final state hadronic observables provide a real tomographic probe of the medium~\cite{tomo}.  
%Measurement of $\gamma$-jet provides new insights in the interaction of quarks
%and gluons with the QGP since
%photon does not interact with the colored medium, which provides 
%the least biased measurement of the energy  and direction
%of the recoiling jet, the essential parts in the study of
%the jet fragmentation modification due to the medium. 

Several algorithms~\cite{gammajetGustavo} have been developed to identify $\gamma$-jet events in p--p  and Pb--Pb collisions,
demonstrating the feasibility of such measurements 
with the ALICE detectors.  However, the jet identification 
%photon tagging jet measurements still require to reconstruct the jet of hadrons emitted opposite to the photon direction, which
remains challenging in the heavy-ion environment in particular for the energies $E_{\gamma} \sim~30$~GeV where $\gamma$-jet events are measurable in ALICE with sufficient statistics. An equivalent approach is to 
%This is not the case if one 
measure  direct-photon--hadrons correlation~\cite{corr}. 
%In contrast to single or di-hadron measurements where the initial parton energy is not known, jets recoiling from direct photon exactly balance the photon's momentum at the leading order (LO). This allows for the determination of jet fragmentation function without full jet reconstruction. Moreover, hadronic observables are strongly biased towards surface production, whereas photons sample the entire collision and should therefore favor the core of the overlap zone where the nucleon density is greatest, which make $\gamma$-h correlation measurements have greater sensitivity to the path-length dependence.  
%Main processes contributing to direct photon production are Compton scattering (q + g $ \rightarrow  \gamma $+q) and quark annihilation (q+$ \bar{q}  \rightarrow  \gamma$+g), beyond the LO, the components from bremsstrahlung and initial state radiation will distort the balance between the photon and hadron, however, by applying an isolation cut, these contributions will be largely suppressed. 

In the following, 
%we discuss the QGP tomography with photon tagged jets in high energy nuclear collisions.
we have first established the intrinsic properties ($k_{T}$) of $\gamma$-jet events expected in pp collisions at LHC energies.
%predicted by the PYTHIA~\cite{pythia} event generator. 
Then we
discuss the nucleus-nucleus (AA) collision case, in particular, we explore the possibility to select $\gamma$-jet events as a function of their localization in the medium to validate the tomographic approach.
%according to the location of the hard process in the medium or
%equivalently to the distance along which the jet travels in the
%medium.
%hard processes occouring at the surface of the medium can
%be triggered by hadrons with large $x_{E}$ values, while hadrons
%with small $x_{E}$ values select hard processes taking place in
%the bulk of the medium. 

\section{$\gamma$-hadron topololigy in pp collisions}
At leading order perturbative QCD, a pair of hard-scattered partons emerges exactly
back-to-back in the center of mass of the partonic system. Due to the finite size of the proton, however, it
was found that each of the colliding parton carries initial
transverse momentum with respect to the colliding axis, originally described as "intrinsic $k_{T}$". Beyond the leading order, initial and final state radiations (ISR/FSR) will generate additional transverse momentum. 
Therefore, the resulting total transverse momentum of the outgoing parton
pair causes an acoplanar and a momentum imbalance,
$<k_{T}>$~\cite{kt}. It is measured as the net transverse momentum of the outgoing
parton-pair $<p_{T}>_{pair}~=~\sqrt{2}\cdot<k_{T}>$. It is anticipated that
medium effects will generate additional transverse momentum resulting in a 
broadening of $k_{T}$ . 
This transverse momentum broadening can be directly related to the transport
parameter $\hat{q}$, which describes the
transverse momentum transferred from  the medium to the traversing parton~\cite{qhat}. 

Using the PYTHIA event generator~\cite{pythia}, we have established the collision energy dependence of $<k_{T}>$ from
%and taking into account the initial and final state radiation (ISR/FSR)
%additionally, within $\gamma$ and jet pair from PYTHIA generated
$\gamma$-jet events, by taking available data 
%we reproduce few data points of the measured $k_{T}$ 
from different experiments measurements~\cite{kTMea}
and extrapolate to the LHC energies. The dependence is
%by fitting on these data points with the function
$<p_{T}>_{pair}=A\cdot\log(B\cdot\sqrt{s})$ with $A=2.064\pm 0.171$ and $B=0.164\pm 0.045$. 

To study the dependence of $<k_{T}>$ with the
transverse momentum of the hard scattering, we have generated $\gamma$-jet and jet-jet
events with PYTHIA generator in different $p_{T}$ bins with collision energy 14~TeV, within
$k_{T}$ setting predicted above and ISR/FSR on.
%A Landau function is used to do the
%fitting on $p_{T}^{pair}$ distribution to get the averaged value
%$<p_{T}>_{pair}$. 
The  averaged $<p_{T}>_{pair}$ versus the transverse momentum,  shows a weakly linear dependence.

\section{Medium modification by heavy ion collisions}

The tomography measurement can be performed by 
%goal of the $\gamma$-h correlation is to preform tomographic studies of the Quark-Gluon Plasma by measuring the path-length dependence of parton energy loss. As explained earlier, the distribution of hard
%scattering vertices sampled by direct photon-hadrons
%correlation is unbiased by the trigger condition.
%Suppression of the opposite jet is averaged over all path-lengths
%given by the distribution of hard scattering vertices. By
selecting $\gamma$-h pairs with different values of the parameter $x_{E}$ =
-$\vec{p}_{T}^{h}\cdot \vec{p}_{T}^{\gamma}/\mid p_{T}^{\gamma}\mid
^{2}$.  This criteria
can effectively control hadron emission from different regions of
the medium and therefore extract the corresponding jet
modification parameters~\cite{tomo}. 
%In this way, the average
%path-length of the away-side parton may then be varied in a well
%controlled manner by selecting events of various momentum
%differences between the $\gamma$-h pair. 

To simulate the medium induced energy loss, we used the Monte-Carlo model 
%is needed for pp and AA simulation, since PYTHIA can
%only be used for pp collision, a HIJING~\cite{hijing} generator is taken for
%heavy ion collision. A simple quenching model (QPYTHIA),  
QPYTHIA~\cite{qpythia}, which combines an energy
loss mechanism~\cite{XN} and a realistic description of the collision
geometry~\cite{glauber}. The HIJING ~\cite{hijing} generator was used to simulate the underlying events of heavy-ion collisions and PYTHIA to simulate pp collisions. 
%developed by N. Armesto et al~\cite{qpythia} is borrowed for studying medium
%induced energy loss~\cite{XN},
Three
samples of $\gamma$-jet events were generated with photon energy larger than 20~GeV. A first sample of pp collisions at 5.5~TeV generated with PYTHIA provides the baseline. The second sample consists in similar events modified by QPYTHIA merged with central collision events . The last sample is obtained by merging the PYTHIA events and peripheral collision events. 
%\begin{itemize}
%\item A first sample is composed of events generated in pp
%collisions at 5.5~TeV, using PYTHIA under AliRoot framework~\cite{aliroot},
%forcusing on the $\gamma$-jet production. No quenching is
%considered. This sample is representative as a baseline of
%quenching study. \item A second sample, which is a merging of the
%quenched $\gamma$-jet events by QPYTHIA, with a background from
%heavy ion collisions,obtained with HIJING computations for most
%central collisions. \item A last sample was generated with the
%same parameters, but the events are not quenched, which merged
%into a heavy ion background from HIJING for non-central
%collisions.
%\end{itemize}

Tagging events with a direct photon well identified~\cite{photon} by the ALICE calorimeters and
measuring the distribution of hadrons emitted oppositely to the
photon as a function of $x_{E}$,  allows us to determinate the jet fragmentation function~\cite{corr}. The underlying event is subtracted by correlating the isolated photon with charged hadrons emitted on the same side as the photon, in the azimuthal range $-\pi/2 < \Delta \phi < \pi/2$. To quantify the medium modification,
%of $\gamma$-hadrons correlation
%distribution in heavy ion collisions relative to pp collisions,
 $I_{ AA}$ is calculated (Fig.~\ref{fig:Iaa}) 
\begin{eqnarray}
\label{eq}
I_{ AA}(x_{E})=\frac{CF_{AA}}{CF_{pp}}\; 
\end{eqnarray}
as the ratio of $\gamma$-hadrons correlation distribution measured in AA and pp collisions. The expected 
enhancement at low $x_{E}$ and suppression at high $x_{E}$ for central collision is observed, whereas, $I_{ AA}$ is equal to 1 for peripheral collisions, where quenching effects are absent. 

\begin{center}
\includegraphics[width=8cm]{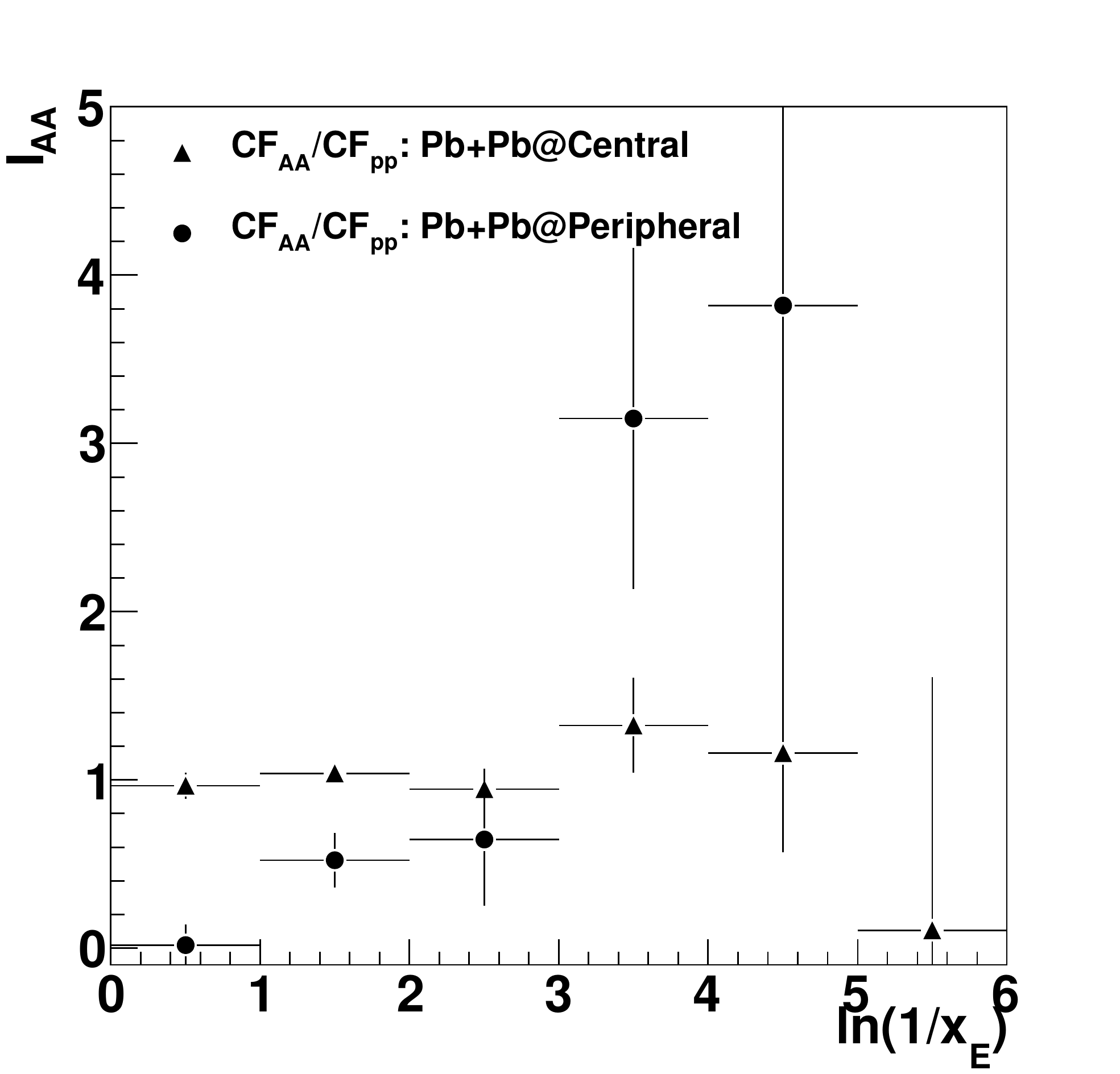}
\figcaption{\label{fig:Iaa}  \em The nuclear modification factor $I_{ AA}$ for
$\gamma$-hadrons correlation distribution in central and peripheral Pb+Pb collisions at $\sqrt{s_{NN}}=~$5.5~TeV.}
\end{center}

To illustrate the selectivity of the tomographic measurement,  the length, L, 
%picture of volume and surface emission,  the length of 
the jet travels inside the
medium is calculated. Fig.~\ref{fig:Qlch} indicates that most high $p_{T}$ leading particles are preferentially produced at the surface (small L),
while low $p_{T}$ leading particles are produced inside the whole volume (large L), which demonstrates the L dependence of of the $\gamma$ tagged charged hadron production for 2 different $x_{E}$ regions. 
\begin{center}
\includegraphics[width=8cm]{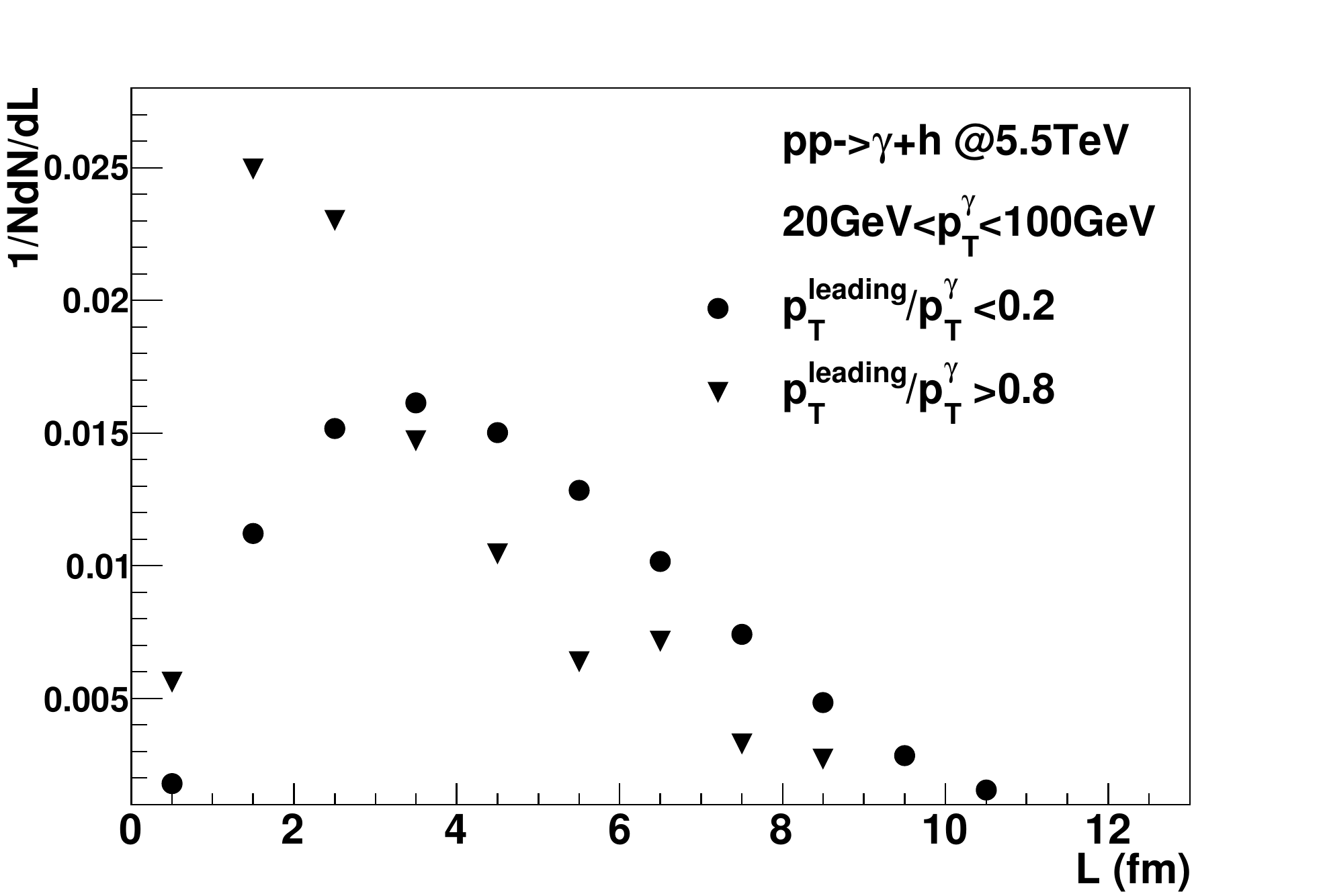}
\figcaption{\label{fig:Qlch}  \em The probability of the leading particles production
as a function of medium length L. }
\end{center}

%according to its kinematics information and
%the nuclear geometry created by QPYTHIA model. Since for $\gamma$-hadrons correlation
%measurement, a full jet reconstruction will never be done, 
%at the opposite hemispherer, is searched from each $\gamma$-jet event, and medium
%length dependence of leading particles are studied, 
%We can interpret the high $p_{T}$ leading particles are more preferentially produced at the surface of the medium (with a small L),
%while low $p_{T}$ leading particles in the whole volume, which demonstrates that large
%$x=p_{T}^{leading}/p_{T}^{\gamma}$ particles are indeed mostly from the surface. 
%No L dependence
%will be found if no quenching is taken into account since no real nuclear geometry is taken actually. %In addition, the
%quenching effect on high $p_{T}$ leading particles and low $p_{T}$
%could be seen on Fig.~\ref{fig:rQ}, 

%\begin{center}
%\includegraphics[width=4cm]{rQch.eps}
%\figcaption{\label{fig:rQ}  \em The ratio with quenching over without quenching for
%leading particles with large $x$ and small $x$ as a function of
%medium length L.}
%\end{center}

We have then studied the L dependence of the medium modification factor $I_{AA}$  (Fig.~\ref{fig:IaaL}) by selecting different $x_{E}$ regions. 
%show different behaviors
%of medium modification factor $I_{AA}$ depending on the medium
%length L (Fig.~\ref{fig:IaaL}), 
For large $x_{E}$ particles, an obvious suppression is observed, 
and the suppression is
stronger with increasing the medium length. For small $x_{E}$, the opposite behavior is obtained as an enhancement ($I_{AA}>$~1). This result implies that
$\gamma$-hadrons correlation could be used to probe volume versus surface emission by selecting $\gamma$-jet events with different $x_{E}$ values.  However such L
dependence will be challenging to measure in the experiments. 

\begin{center}
\includegraphics[width=8cm]{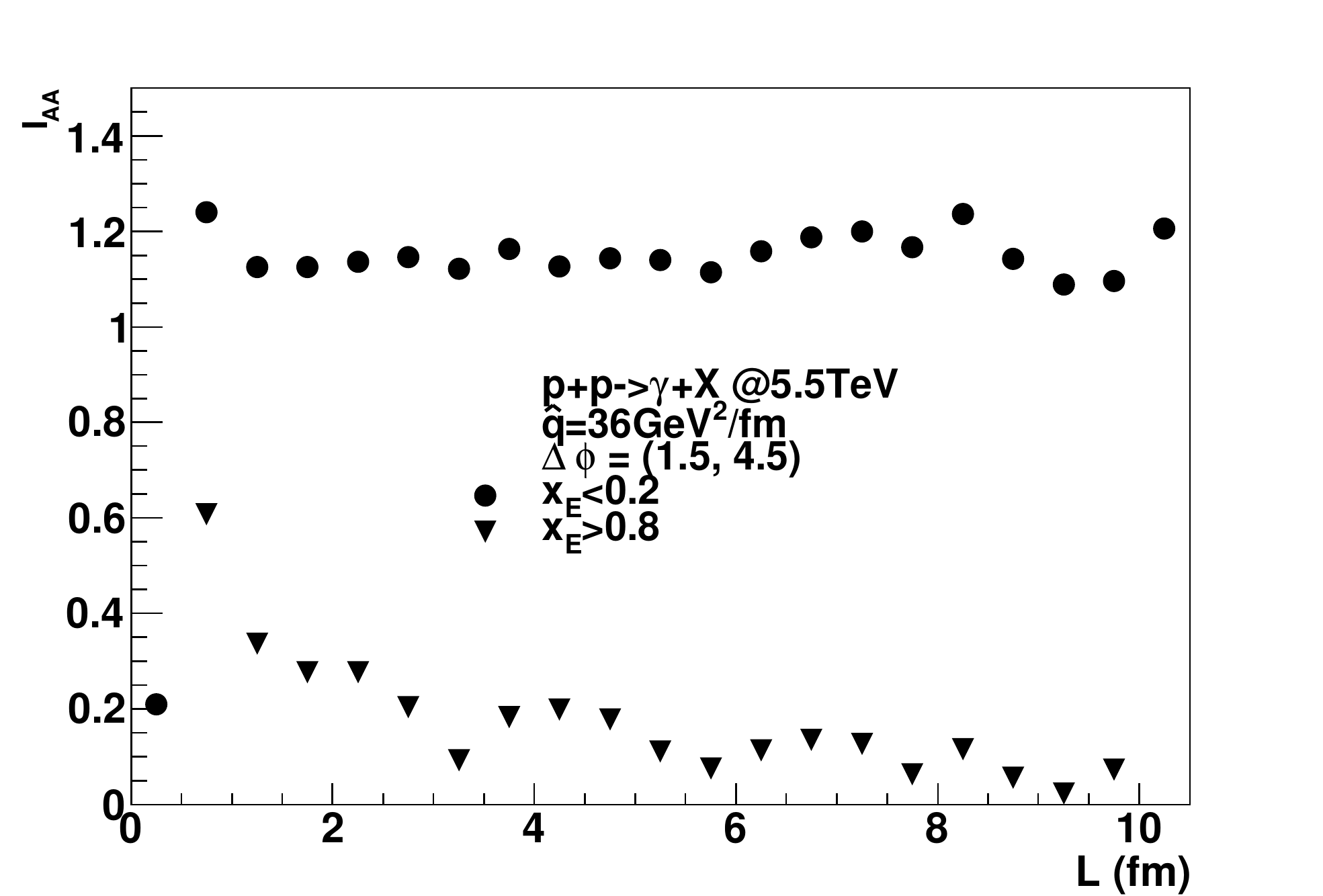}
\figcaption{\label{fig:IaaL}  \em The nuclear modification factor $I_{AA}$ distribution
as a function of medium length L by selecting different regions of
$x_{E}$ on correlation distribution.}
\end{center}

\section{Conclusions}
$\gamma$+jet studies are widely recognized as a powerfull tool to characterize QGP. The "$\gamma$+jet tomography" study will enable us to extract jet quenching
parameters in different regions of the dense medium via
measurement of the nuclear modification factor of $\gamma$-hadrons
correlation. 
%The research presented here represents significant progress towards achieving one of the primary goals at the LHC program: to determine the density profile of the Quark-Gluon Plasma created in heavy-ion collisions by measing its interaction with, and response to, jets.

\acknowledgments{We especially thank  Prof.Xin-Nian Wang, Prof.Andreas Morsh, Prof.Peter Jacobs and Dr.Yuri Kharlov for their enthustic and fruitful discussions, also the full PWG4 workgroup in ALICE collabration.}

\end{multicols}

\end{document}